\documentclass[11pt,aps,preprint]{revtex4}
\usepackage[active]{srcltx}
\usepackage[utf8]{inputenc}
\usepackage{latexsym}
\usepackage{amsmath}
\usepackage{graphicx}

\begin{document}

\title{Supersymmetric Extension of the Quantum Spherical Model}

\author{Pedro R. S. Gomes}
\email{pedrorsg@fma.if.usp.br}
\affiliation{Instituto de F\'\i sica, Universidade de S\~ao Paulo\\
Caixa Postal 66318, 05315-970, S\~ao Paulo, SP, Brazil}%

\author{P. F. Bienzobaz}
\email{paulafb@if.usp.br}
\affiliation{Instituto de F\'\i sica, Universidade de S\~ao Paulo\\
Caixa Postal 66318, 05315-970, S\~ao Paulo, SP, Brazil}%

\author{M. Gomes}
\email{mgomes@fma.if.usp.br}
\affiliation{Instituto de F\'\i sica, Universidade de S\~ao Paulo\\
Caixa Postal 66318, 05315-970, S\~ao Paulo, SP, Brazil}%


\begin{abstract}

In this work, we present a supersymmetric extension of the quantum spherical model, both in components and also
in the superspace formalisms. We find the solution for short/long range interactions through the imaginary time formalism path integral
approach. The existence of critical points (classical and quantum)  is analyzed and the corresponding critical dimensions are determined.

\end{abstract}
\maketitle


\section{Introduction}
It is a  common fact that the methods and ideas of a given area of theoretical physics may be  useful
in another completely distinct context. Besides the technical utility that this sharing of knowledge may offer, it provides
different views of the problem yielding profound implications. A classical example of this sort of situation is the junction of ideas coming from Gell-Mann-Low quantum electrodynamics  and Kadanoff's  block-spin ultimately leading to the renormalization group, useful both in quantum field theory and statistical mechanics \cite{Wilson}.
Proceeding along these lines of thinking, in this paper we  consider the supersymmetric extension of a quantum version of a
traditional model in statistical mechanics, namely, the spherical model \cite{Berlin}.
Supersymmetry, by nontrivially combining internal and
space-time symmetries in a way impossible in the traditional Lie algebraic approach,  has produced new relevant insights in   the context of quantum field theory \cite{Weinberg} and even in  condensed matter physics \cite{Parisi,Sourlas,Junker}. In the nonrelativistic context, supersymmetry was introduced by Nicolai \cite{Nicolai1,Nicolai2} and applied to the study of spin systems.

In quantum mechanics, supersymmetry requires the existence of
supercharges $Q$ and $\bar Q$, such that $\{Q,\bar Q\}=\mathcal H$, where $\mathcal H$ is the Hamiltonian of the system, and furthermore satisfy $[Q,\mathcal H]=[\bar Q,\mathcal H]=0$; these supercharges realize the transmutation of bosonic states into fermionic ones and vice-versa so that the ground state is left invariant. Thus, a supersymmetric theory is characterized by a bosonic ground state $|0\rangle_B$ with energy equal to zero,
i.e., annihilated by the supercharges and by the Hamiltonian
\begin{equation}
Q |0\rangle_B=\bar Q |0\rangle_B=\mathcal H|0\rangle_B=0.
\end{equation}
On the other hand, if the supersymmetry is broken, then, at least one of the supercharges does not annihilate the ground state any
longer. Instead, there is a pair of degenerated ground states bosonic and fermionic, $|0\rangle_B$ and  $|0\rangle_F$, with energy $E_0$, such that
\begin{equation}
|0\rangle_F\equiv \frac{1}{\sqrt{E_0}}\,\bar Q |0\rangle_B~~~\text{and}~~~|0\rangle_B\equiv \frac{1}{\sqrt{E_0}}\, Q |0\rangle_F.
\end{equation}

In contrast with phase transitions at finite temperature, driven by thermal fluctuations and
denoted as classical phase transition, a quantum phase transition occurs at zero temperature and
is caused by quantum fluctuations connected with  Heisenberg's uncertainty relations. Thus,
at zero temperature, a phase transition may occur in a broken supersymmetric situation  when  the ground state energy $E_0$  is non-vanishing;
it is characterized  by some non-thermal coupling parameter $g$, that assumes the value $g_c$ at critical point.
So, near the critical point,
\begin{equation}
E_0\sim |g-g_c|^{z\nu},
\end{equation}
where $z$ and $\nu$ are the dynamical and correlation length critical exponents, respectively.
This suggests that in a situation where supersymmetry holds, the system will not exhibit a critical point because
of the vanishing of the ground state energy independently of the value of $g$. On the contrary, it may display a critical
behavior if supersymmetry is broken. It should be noticed that,
essentially because the differences in the bosonic and fermionic distribution functions, supersymmetry is always broken at finite temperature \cite{Das,Matsumoto,Das1}.

The classical spherical model was initially proposed by Berlin and Kac \cite{Berlin}, as
a simplified continuous version of the Ising model. Since it is exactly soluble, it has been used
to study the critical behavior in a variety of situations \cite{Joyce,Kalok,Jones,Jagannathan,Hornreich,Vojta4,Vojta5,Zohar}.
Typically, the classical solution exhibits an anomaly at low temperature, concerning
to the third law of the thermodynamics. It was suggested that such pathological behavior
could be corrected by introducing quantum fluctuations, which were not taken into account in the classical case \cite{Nieuwenhuizen}.
The extension to the quantum domains has been considered by various authors and has raised much attention in the
context of quantum phase transitions \cite{Hertz,Vojta2,Sachdev}. In particular, we mention studies in some quantum versions of the
spherical model \cite{Nieuwenhuizen,Obermair,Henkel1,Henkel2,Vojta1,Nieuwenhuizen1,Coutinho}, and also including some ingredients of the statistical
mechanics, such as the influence of random fields \cite{Vojta3}, spin glasses \cite{Singh,N1,Kopec,Alba,Alba1}, frustration \cite{Zohar1},
competing interactions \cite{Bienzobaz}, and quantum Lifshitz point \cite{Dutta}.
In this work we extend these studies by considering a supersymmetric version of the spherical model with especial attention to the existence of critical points and the determination of critical dimensions. The supersymmetric constraints are implemented through delta functions in
the partition function and by the saddle point evaluation we obtained the conditions to which the action is stationary.
The critical behavior analysis follows from the study of these equations in some simple cases, according to the values of the
saddle point parameters. For a certain situation, where the bosonic and fermionic degrees of freedom are decoupled,
we found that the critical behavior of the supersymmetrical model reduces to that of the quantum spherical model. On the other hand, when the bosonic and fermionic degrees of freedom are coupled, corresponding to another choice of the saddle point parameters, the model exhibits a distinct critical behavior. We also discussed some issues in connection with the supersymmetry breaking.

Our work is organized as follows. In the section \ref{section3}, we review the bosonic quantum spherical model and
evaluate the partition function through the imaginary time formalism.
In the section \ref{se}, we present the supersymmetric extension, both in components and
also in the superspace formalisms. The partition function of the
supersymmetric model is evaluated in the section \ref{itf} and its critical behavior is discussed in section
\ref{cb}. A summary and additional remarks are presented in the Conclusions. One appendix to study the canonical quantization of the model
is included.


\section{The Quantum Spherical Model}\label{section3}

In this section, we discuss the quantum spherical model subject to the strict spherical
constraint, corresponding to a canonical ensemble.
The quantum version of the model, by implementing the mean spherical constraint, was studied by Vojta \cite{Vojta1}. The classical Hamiltonian of the spherical model is given by
\begin{equation}
\mathcal{H}_{c}=\frac12\sum_{{\bf r},{\bf r}^{\prime}}J_{{\bf r},{\bf r}^{\prime}}S_{{\bf r}}S_{{\bf r}^{\prime}}+h\sum_{{\bf r}}S_{\bf r},
\label{1.1}
\end{equation}
where ${\bf r}$ and ${\bf r}'$ are lattice vectors, $\left\{S_{\bf r}\right\}$ is a
set of spins variables that can assume continuous values, $-\infty<S_{\bf r}<\infty$, in a D-dimensional hypercubic
lattice; $J_{{\bf r},{\bf r}^{\prime}}$ is the interaction energy that depends only on the distance between the sites ${\bf r}$ and ${\bf r}'$, $J_{{\bf r},{\bf r}'}\equiv J(|{\bf r}-{\bf r}'|)$, and $h$ is the external field.
We assume that the $S_{\bf r}$ variables are subject to the spherical constraint
\begin{equation}
\sum_{\bf r} S_{\bf r}^2=N
\label{1.2},
\end{equation}
where $N$ is the total number of lattice sites. On the other hand, the mean spherical constraint is defined as $\sum_r\left<S_{\bf r}^2\right>=N$, where $\left<\cdots\right>$ designates a thermal average.  Of course, in the thermodynamic limit, $N\rightarrow\infty$, these
two constraints yield the same results, as we shall see shortly.
A discussion about the properties of spherical and mean spherical models in the classical context is found in \cite{Joyce}.

In order to construct the quantum model we need to add to the Hamiltonian a  kinetic term involving the
conjugate momentum variable to $S_{\bf r}$, denoted by $P_{\bf r}$, and then promote such variables to  operators
satisfying the usual canonical commutation relations
\begin{equation}
[S_{\bf r}, S_{{\bf r}^{\prime}}]=0, ~~~[P_{\bf r}, P_{{\bf r}^{\prime}}]=0~~~\text{and}~~~
[S_{\bf r}, P_{{\bf r}^{\prime}}]=i\delta_{{\bf r}, {\bf r}^{\prime}}.
\end{equation}
Including then a kinetic term quadratic in the momenta, the quantum Hamiltonian becomes
\begin{equation}
\mathcal{H}=\frac12 g \sum_{\bf r} P_{\bf r}^2
+\frac12\sum_{{\bf r},{\bf r}^{\prime}}J_{{\bf r},{\bf r}^{\prime}}S_{{\bf r}}S_{{\bf r}^{\prime}}+h\sum_{{\bf r}}S_{\bf r}.
\end{equation}
The coupling constant $g$ measures the relevance of the quantum fluctuations and the limit $g\rightarrow 0$
corresponds to the classical regime.
To construct the supersymmetric version it is convenient to use the Lagrangian formulation. Moreover, it is also useful for the evaluation of the quantum partition function through the imaginary time formalism, as we will explicitly show in the next subsection. After a Legendre transformation, we obtain the Lagrangian
\begin{equation}
\mathcal{L}=\frac{1}{2g} \sum_{\bf r} \dot{S}_{\bf r}^2
-\frac12\sum_{{\bf r},{\bf r}^{\prime}}J_{{\bf r},{\bf r}^{\prime}}S_{{\bf r}}S_{{\bf r}^{\prime}}-h\sum_{{\bf r}}S_{\bf r},
\end{equation}
where the dot means a derivative with respect to the time. For simplicity,
in the remaining of this paper, we will  take the external field equal to zero, $h=0$. It must be clear that we can consider the
external field dependence without difficulties.


\subsection{Imaginary Time Formalism}

Let us evaluate the partition function through the path integral imaginary time formalism approach \cite{Abrikosov}.
Thus, we need to pass to the Euclidean imaginary time $\tau=it$, with $\tau\in[0,\beta]$ and
$\beta$ the inverse of the temperature. Furthermore, the bosonic variables are
required to satisfy the periodic boundary condition $S_{\bf r}(0)=S_{\bf r}(\beta)$, which
gives rise to the discrete spectrum Matsubara frequencies $\omega_n=2n\pi/\beta$, with $n\in \text{Z}$.
The partition function is given by
\begin{equation}
Z=\int \mathcal{D}S_{\bf r}\,\delta\Big{(}\sum_{\bf r}S_{\bf r}^2-N\Big{)}e^{-\int_{0}^{\beta}d\tau\mathcal{L}_E},
\end{equation}
where the Euclidean Lagrangian is
\begin{equation}
\mathcal{L}_E=\frac{1}{2g} \sum_{\bf r}\left( \frac{\partial S_{\bf r}(\tau)}{\partial\tau}\right)^2
+\frac12\sum_{{\bf r},{\bf r}^{\prime}}J_{{\bf r},{\bf r}^{\prime}}S_{{\bf r}}S_{{\bf r}^{\prime}}.
\end{equation}
The integration measure $\mathcal{D}S_{\bf r}$ symbolically stands for functional integration over the spins variable of all sites of the lattice,
i.e., $\mathcal{D}S_{\bf r}\equiv \prod_{\bf r} \mathcal{D}S_{\bf r}$.
Employing the functional integral representation for the delta function
\begin{equation}
\delta\big{(}\sum_{\bf r}S_{\bf r}^2-N\big{)}=\int\mathcal{D} \lambda\, e^{{-\int_{0}^{\beta}}d\tau\lambda\big{(}\sum_{\bf r}S_{\bf r}^2-N\big{)}},
\end{equation}
we can write
\begin{equation}
Z=\int \mathcal{D}S_{\bf r}\mathcal{D}\lambda\, \exp\left\{-\int_{0}^{\beta}d\tau\left[\frac{1}{2g} \sum_{\bf r}\left( \frac{\partial S_{\bf r}(\tau)}{\partial\tau}\right)^2
+\frac12\sum_{{\bf r},{\bf r}^{\prime}}J_{{\bf r},{\bf r}^{\prime}}S_{{\bf r}}S_{{\bf r}^{\prime}}+
\lambda\sum_{\bf r}S_{\bf r}^2-\lambda N\right]\right\}.
\end{equation}
This is an appropriate representation, since the integral on $S_{\bf r}$ becomes Gaussian and can be integrated out.
Before continuing, however, it is convenient to introduce the Fourier transformation of $S_{\bf r}$
\begin{equation}
S_{\bf r}(\tau)=\frac{1}{\sqrt{N}}\sum_{\bf q}e^{i {\bf q}\cdot {\bf r}}S_{\bf q}(\tau).
\end{equation}
With this, the Euclidean action acquires a simple form and the partition function becomes
\begin{equation}
Z=\int \mathcal{D}S_{\bf q}\mathcal{D}\lambda\, \exp\left\{-\int_{0}^{\beta}d\tau
\left[\sum_{\bf q}S_{\bf q}\left(-\frac{1}{2g}\frac{\partial^2}{\partial \tau^2}+\frac12J({\bf q})+\lambda\right)S_{-{\bf q}}
-\lambda N\right]\right\}.
\end{equation}
Here, we have identified $J({\bf q})$ as the Fourier transformation of the interaction energy $J_{{\bf r},{\bf r}^{\prime}}$,
\begin{equation}
J({\bf q})=\sum_{\bf h}J(|{\bf h}|)e^{i {\bf q}\cdot{\bf h}},~~~\text{with}~~~{\bf h}={\bf r}-{\bf r}^{\prime}.
\end{equation}
After the Gaussian integration and using the identity $\det A=e^{\text{Tr} \ln A}$, we get
\begin{equation}
Z=\int \mathcal{D}\lambda\, e^{- N S_{eff}},
\end{equation}
where we defined
\begin{equation}
S_{eff}\equiv -\int_{0}^{\beta}d\tau\, \lambda+
\frac{1}{2N}\,\text{Tr}\left[\sum_{\bf q}\ln \left(-\frac{1}{2g}\frac{\partial^2}{\partial\tau^2}+\frac12J({\bf q})+\lambda \right)\right].
\end{equation}
In the thermodynamic limit, $N\rightarrow\infty$, we may use the saddle point functional method to evaluate
the partition function. The saddle point condition corresponds to the effective action above be stationary:
\begin{equation}
\frac{\delta S_{eff}}{\delta \lambda(\tau)}=0.
\end{equation}
We will suppose that the saddle point $\lambda$ is time independent, and defined as $\lambda\equiv\mu$.
By means of the identity $\delta \text{Tr}\ln A=\text{Tr}A^{-1}\delta A$, we can obtain
\begin{equation}
1-\frac{1}{2N\beta}\sum_{\bf q}\sum_{n=-\infty}^{\infty}\frac{1}{\frac{\omega_n^2}{2g}+\frac{1}{2}J({\bf q})+\mu}=0.
\end{equation}
The sum on integers can be calculated thanks to the identity
\begin{equation}
\sum_{n=-\infty}^{\infty}\frac{1}{n^2+y^2}=\frac{\pi}{y}\,\coth(\pi y),~~~y>0.
\end{equation}
The final result is
\begin{equation}
1-\frac{1}{N}\sum_{\bf q}\frac{g}{2\omega_{\bf q}}\coth\left(\frac{\beta\omega_{\bf q}}{2}\right)=0,
\label{v1}
\end{equation}
where, $\omega_{\bf q}^2\equiv 2g(\mu+J({\bf q})/2)$. This result is exactly that obtained with
the mean spherical constraint \cite{Vojta1}. From this expression, we can determine the critical properties
of the model at finite temperature as well as at $T=0$. The analysis of the critical behavior can be done
by considering the system near the critical point, with $\mu\rightarrow 0$ and the interaction parameterized
as $J({\bf q})\sim q^x$, $q\equiv |{\bf q}|$, for small momenta. The parameter $x$ determines the short or long range character of the interaction.
Typically, for short range interactions we have $x=2$. Despite being described quantum mechanically, for any finite temperature
the system shows a classical critical behavior similar that obtained from the classical version of the model.
Essentially, this is because the thermal fluctuations generally dominate quantum fluctuations at macroscopic scales. At zero temperature,
however, there is a quantum phase transition
characterized  by new critical exponents \cite{Vojta1}.

\section{Supersymmetric Extension}\label{se}

In this section, we shall construct the supersymmetric version of the
spherical model in terms of the component fields formulation as well as in the superspace.
The fundamental ingredient in this extension is the introduction of additional fermionic degrees of
freedom at each lattice site to balance the bosonic ones. Namely, at each site of the lattice, besides
the bosonic variable $S_{\bf r}$, we will associate the fermionic counterparts $\psi_{\bf r}$ and $\bar\psi_{\bf r}$
($\bar\psi_{\bf r}\equiv\psi_{\bf r}^{\dagger}$), that
in the quantum case satisfy the anti-commutation relations
\begin{equation}
\{\psi_{\bf r},\psi_{{\bf r}^{\prime}}\}=0,~~~\{\bar\psi_{\bf r},\bar\psi_{{\bf r}^{\prime}}\}=0~~~\text{and}~~~
\{\psi_{\bf r},\bar\psi_{{\bf r}^{\prime}}\}=\delta_{{\bf r},{\bf r}^{\prime}}.
\end{equation}

\subsection{Components Formulation}

The natural form for the supersymmetric Lagrangian is
\begin{equation}
\mathcal{L}_{Susy}=\frac{1}{2g} \sum_{\bf r} \dot{S}_{\bf r}^2
-\frac12\sum_{{\bf r},{\bf r}^{\prime}}J_{{\bf r},{\bf r}^{\prime}}S_{{\bf r}}S_{{\bf r}^{\prime}}+
\frac{i}{\sqrt{g}}\sum_{\bf r}\bar{\psi}_{\bf r}\dot{\psi}_{\bf r}-
\sum_{{\bf r},{\bf r}^{\prime}}U_{{\bf r},{\bf r}^{\prime}}\bar{\psi}_{{\bf r}}\psi_{{\bf r}^{\prime}}.
\label{c0}
\end{equation}
Of course, the requirement of invariance under supersymmetry transformations will imply a relation between
the interactions $J_{{\bf r},{\bf r}^{\prime}}$ and $U_{{\bf r},{\bf r}^{\prime}}$. We are supposing that the interaction
$U_{{\bf r},{\bf r}^{\prime}}$ also depends only on the distance between the sites ${\bf r}$ and ${\bf r}^{\prime}$.
In addition, the constraint on the bosonic variable will imply others constraints involving the fermionic variables, which will
be discussed shortly.

Now, it is easy to verify that the set of transformations
\begin{equation}
\delta_{\epsilon}S_{\bf r}=\bar{\psi}_{\bf r}\epsilon,~~~
\delta_{\epsilon}\psi_{\bf r}=-\frac{i}{\sqrt{g}}\,\dot{S}_{\bf r}\epsilon-
\sum_{{\bf r}^{\prime}}U_{{\bf r},{\bf r}^{\prime}}S_{{\bf r}^{\prime}}\epsilon~~~\text{and}~~~
\delta_{\epsilon}\bar{\psi}_{\bf r}=0
\label{c1}
\end{equation}
and
\begin{equation}
\delta_{\bar\epsilon}S_{\bf r}=\bar{\epsilon}{\psi}_{\bf r},~~~\delta_{\bar\epsilon}\psi_{\bf r}=0
~~~\text{and}~~~ \delta_{\bar\epsilon} \bar\psi_{\bf r}=\frac{i}{\sqrt{g}}\,\dot{S}_{\bf r}\bar\epsilon-
\sum_{{\bf r}^{\prime}}U_{{\bf r},{\bf r}^{\prime}}S_{{\bf r}^{\prime}}\bar\epsilon
\label{c2}
\end{equation}
leave the Lagrangian (\ref{c0}) invariant up to surface terms, i.e., up to a total derivative, provided that
\begin{equation}
\sum_{\bf s}U_{{\bf r},{\bf s}}U_{{\bf s},{\bf r}^{\prime}}\equiv J_{{\bf r},{\bf r}^{\prime}}.
\end{equation}
Notice that the Fourier transformation of this relation furnishes $[U({\bf q})]^2=J({\bf q})$.
The parameters of transformations $\epsilon$ and $\bar{\epsilon}$ are anticommuting infinitesimal quantities.
Equations (\ref{c1}) and (\ref{c2}) are the supersymmetry transformations, that relate the bosonic degrees
of freedom to the fermionic ones.

The next step is to investigate the consistency of the constraints. More precisely, the implications of the supersymmetry
transformations on the spherical constraint
\begin{equation}
\sum_{\bf r} S_{\bf r}^2=N.
\label{c3}
\end{equation}
It can be verified that, under the transformations (\ref{c1}) and (\ref{c2}), we are led to the additional constraints
\begin{equation}
\sum_{\bf r}\bar{\psi}_{\bf r}S_{\bf r}=0,~~~\sum_{\bf r}\psi_{\bf r}S_{\bf r}=0~~~\text{and}~~~
\sum_{\bf r}\bar{\psi}_{\bf r}{\psi}_{\bf r}=-\sum_{{\bf r},{\bf r}^{\prime}}U_{{\bf r},{\bf r}^{\prime}}S_{\bf r}S_{{\bf r}^{\prime}}.
\label{c4}
\end{equation}
In the last relation, we have discarded a surface term.
In summary, to have a consistent supersymmetric formulation of the spherical model we need now  four constraints.
These constraints introduce effectively an interaction between the bosonic and fermionic variables.
They can be implemented via four Lagrange multipliers, being two of bosonic character and two of fermionic character.

\subsection{Superspace Formulation}\label{ssb}

A very elegant and concise way to formulate supersymmetry is through the notion of the superspace.
Moreover, this formulation has the advantage of making the underlying theory manifestly supersymmetric.
The price is that it is necessary to introduce auxiliary (not dynamical) bosonic degrees of freedom.
The superspace consists of an extension of the ordinary space, that in the quantum mechanic case corresponds to the time coordinate only,
in order to accommodate the anticommuting coordinates $\theta$ and $\bar\theta$, satisfying $\theta^2=\bar\theta^2=0$ and $\{\theta,\bar\theta\}=0$.
To fix our notation, let us define some useful operations with these Grassmannian variables:
\begin{equation}
\frac{\partial}{\partial \theta} \,\theta =\frac{\partial}{\partial \bar\theta} \,\bar\theta\equiv 1 ,~~~
\frac{\partial}{\partial \bar\theta}\, \theta =\frac{\partial}{\partial \theta}\, \bar\theta\equiv 0
\end{equation}
and
\begin{equation}
\int d\theta=\int d\bar\theta\equiv 0,~~~\int d\theta \,\theta=\int d\bar\theta \,\bar\theta\equiv 1~~~\text{and}~~~
\int d\theta d\bar\theta\,\bar\theta\theta\equiv 1.
\end{equation}
We then associate at each site of the lattice a superfield $\Phi_{\bf r}(t,\theta,\bar\theta)$ that is a function of
the superspace coordinates and can be expanded in powers of $\theta$ and $\bar\theta$ in the following way
\begin{equation}
\Phi_{\bf r}(t,\theta,\bar\theta)\equiv S_{\bf r}+\bar\theta\psi_{\bf r}+\bar\psi_{\bf r}\theta+\bar\theta\theta F_{\bf r}.
\end{equation}
The bosonic and fermionic variables are components of the superfield and $F_{\bf r}$ is the auxiliary field.
Next, let us introduce the supercharges,
\begin{equation}
Q\equiv \frac{\partial}{\partial\bar\theta}+i\theta\frac{\partial}{\partial t}~~~
\text{and}~~~\bar{Q}\equiv -\frac{\partial}{\partial\theta}-i\bar\theta\frac{\partial}{\partial t},
\end{equation}
that are the generators of translations in the superspace,
\begin{equation}
\theta\rightarrow\theta+\epsilon,~~~t\rightarrow t-i\bar\theta\epsilon~~~\text{and}~~~
\bar\theta\rightarrow\bar\theta+\bar\epsilon,~~~t\rightarrow t+i\bar\epsilon\,\theta.
\end{equation}
Also, it is easy to verify that the superfield transformations,
\begin{equation}
\delta_{\epsilon}\Phi_{\bf r}=\bar{Q}\epsilon \Phi_{\bf r}~~~\text{and}~~~\delta_{\bar\epsilon}\Phi_{\bf r}=\bar\epsilon Q\Phi_{\bf r},
\end{equation}
correspond to the following component transformations
\begin{equation}
\delta_{\epsilon}S_{\bf r}=\bar{\psi}_{\bf r}\epsilon,~~~
\delta_{\epsilon}\psi_{\bf r}=-i\dot{S}_{\bf r}\epsilon+
F_{\bf r}\epsilon,~~~
\delta_{\epsilon}\bar{\psi}_{\bf r}=0~~~\text{and}~~~\delta_{\epsilon}F_{\bf r}=i\dot{\bar\psi}_{\bf r}\epsilon
\end{equation}
and
\begin{equation}
\delta_{\bar\epsilon}S_{\bf r}=\bar\epsilon{\psi}_{\bf r},~~~
\delta_{\bar\epsilon}\psi_{\bf r}=0,~~~
\delta_{\bar\epsilon}\bar{\psi}_{\bf r}=i\dot{S}_{\bf r}\bar\epsilon+
F_{\bf r}\bar\epsilon~~~\text{and}~~~\delta_{\epsilon}F_{\bf r}=-i\bar\epsilon\dot{\psi}_{\bf r}.
\end{equation}
These transformations may be compared with (\ref{c1}) and (\ref{c2}) by rescaling the variables $S_{{\bf r}}\rightarrow g^{-1/2}S_{{\bf r}}$, $\psi_{{\bf r}}\rightarrow g^{-1/4}\psi_{{\bf r}}$ and $\bar\psi_{{\bf r}}\rightarrow g^{-1/4}\bar\psi_{{\bf r}}$
and also the parameters  $\epsilon\rightarrow g^{-1/4}\epsilon$ and $\bar\epsilon\rightarrow g^{-1/4}\bar\epsilon$.
Furthermore, the supercharges satisfy the algebra
\begin{equation}
\{Q,Q\}=0,~~~\{\bar{Q},\bar{Q}\}=0~~~\text{and}~~~\{Q,\bar{Q}\}=-2i\frac{\partial}{\partial t}.
\end{equation}
The last anticommutation relation is proportional to the generator of time translations that must be identified with the Hamiltonian.

We need to construct an action in the superspace such that the corresponding Lagrangian reproduces (\ref{c0}) after
integration over $\theta$ and $\bar\theta$, and further take in to account the constraints (\ref{c3}) and (\ref{c4}).
Initially, let us consider the kinetic term. For this purpose it is convenient to introduce the covariant derivatives
\begin{equation}
D\equiv -\frac{\partial}{\partial\bar\theta}+i\theta\frac{\partial}{\partial t}~~~
\text{and}~~~\bar{D}\equiv \frac{\partial}{\partial\theta}-i\bar\theta\frac{\partial}{\partial t},
\end{equation}
that satisfy
\begin{equation}
\{D,Q\}=\{D,\bar{Q}\}=\{\bar{D},Q\}=\{\bar{D},\bar{Q}\}=0~~~\text{and}~~~\{D,\bar{D}\}=2i\frac{\partial}{\partial t}.
\end{equation}
These anticommutation relations imply that the covariant derivative of a superfield has the same property under supersymmetry
transformations as the superfield itself. So, any action involving only superfields as well as covariant derivatives of superfileds is
manifestly supersymmetric.

Now, observe that
\begin{equation}
D\Phi_{\bf r}=-\psi_{\bf r}+\left(i\dot{S}_{\bf r}-F_{\bf r}\right)\theta-i\dot\psi_{\bf r}\bar\theta\theta
\end{equation}
and
\begin{equation}
\bar{D}\Phi_{\bf r}=-\bar\psi_{\bf r}-\left(i\dot{S}_{\bf r}+F_{\bf r}\right)\bar\theta+i
\dot{\bar\psi}_{\bf r}\bar\theta\theta.
\end{equation}
Then, the kinetic term can be obtained according to
\begin{equation}
\frac12 \sum_{\bf r}\bar{D}\Phi_{\bf r}D\Phi_{\bf r}\Big{|}_{\bar\theta\theta}=\frac{1}{2} \sum_{\bf r} \dot{S}_{\bf r}^2+
i\sum_{\bf r}\bar{\psi}_{\bf r}\dot{\psi}_{\bf r}+\frac{1}{2}\sum_{\bf r}F_{\bf r}^2.
\label{kp}
\end{equation}
Notice that we do not have terms involving time derivative of the variable $F_{\bf r}$, which means
that it is an auxiliary variable as we said above, i.e., it does not possess dynamics and can be eliminated via its equation of motion.
The interaction terms can also be constructed in a simple way
\begin{equation}
\frac{1}{2}\sum_{{\bf r},{\bf r}^{\prime}}U_{{\bf r},{\bf r}^{\prime}}\Phi_{\bf r}\Phi_{{\bf r}^{\prime}}\Big{|}_{\bar\theta\theta}=
\sum_{{\bf r},{\bf r}^{\prime}}U_{{\bf r},{\bf r}^{\prime}}S_{\bf r}F_{{\bf r}^{\prime}}-
\sum_{{\bf r},{\bf r}^{\prime}}U_{{\bf r},{\bf r}^{\prime}}\bar\psi_{\bf r}\psi_{{\bf r}^{\prime}}.
\label{ip}
\end{equation}
The interaction energy  $J_{{\bf r},{\bf r}^{\prime}}$ does not appear explicitly in the action, but only after the
elimination of the auxiliary field $F_{\bf r}$.
The action in the superspace is then the sum of the kinetic and the interaction parts
\begin{equation}
S=\int dt d\theta d\bar\theta \left(\frac12 \sum_{\bf r}\bar{D}\Phi_{\bf r}D\Phi_{\bf r}+
\frac{\sqrt{g}}{2}\sum_{{\bf r},{\bf r}^{\prime}}U_{{\bf r},{\bf r}^{\prime}}\Phi_{\bf r}\Phi_{{\bf r}^{\prime}}\right).
\end{equation}
The equation of motion for the auxiliary variable $F_{\bf r}$ is given by
\begin{equation}
F_{\bf r}=-\sum_{{\bf r}^\prime}U_{{\bf r},{\bf r}^{\prime}}S_{{\bf r}^\prime},
\label{eom}
\end{equation}
that can be used to eliminate $F_{\bf r}$. The resulting action of this process is,
up to a rescaling of the fields $S_{{\bf r}}\rightarrow g^{-1/2}S_{{\bf r}}$, $\psi_{{\bf r}}\rightarrow g^{-1/4}\psi_{{\bf r}}$
and $\bar\psi_{{\bf r}}\rightarrow g^{-1/4}\bar\psi_{{\bf r}}$ as mentioned before, just that of components formulation.

The spherical constraint will be imposed on the superfield according to
\begin{equation}
\sum_{\bf r}\Phi_{\bf r}\Phi_{\bf r}=N,
\end{equation}
which in components yields the following relations:
\begin{equation}
\sum_{\bf r} S_{\bf r}^2=N,~~~\sum_{\bf r}\bar{\psi}_{\bf r}S_{\bf r}=0,~~~\sum_{\bf r}\psi_{\bf r}S_{\bf r}=0~~~\text{and}~~~
\sum_{\bf r}\bar{\psi}_{\bf r}{\psi}_{\bf r}=\sum_{\bf r}S_{\bf r}F_{{\bf r}}.
\end{equation}
After using the equation (\ref{eom}), we get (\ref{c3}) and (\ref{c4}).
We can implement the constraint through a delta function (inside the partition function) that in the integral representation
requires an integration over a superfield
\begin{equation}
\Xi\equiv\gamma +\bar\theta \zeta+\bar\zeta\theta+\bar\theta\theta\lambda,
\end{equation}
where $\gamma$ and $\lambda$ play the role of the usual Lagrange multipliers and $\zeta$ and $\bar\zeta$ of anticommuting Lagrange multipliers.
To sum up, we have constructed a consistent supersymmetric extension of the quantum spherical model in components as well as in
the superspace formulation.
The next step is to evaluate the partition function, which will be discussed in the sequel.

\section{Imaginary Time Formalism}\label{itf}

We will employ the imaginary time formalism  to the evaluation of the partition function,
as it has been discussed in the purely bosonic model. Because of the anticommuting character of the Grassmannian quantities,
the fermionic variables are required to satisfy the antiperiodic boundary conditions. In general, the fields must satisfy
\begin{equation}
S_{\bf r}(0)=S_{\bf r}(\beta),~~~\psi_{\bf r}(0)=-\psi_{\bf r}(\beta)~~~\text{and}~~~\bar\psi_{\bf r}(0)=-\bar\psi_{\bf r}(\beta).
\label{kms}
\end{equation}
The reflex of the antiperiodic conditions is the arising of the discrete spectrum fermionic frequencies $\omega_n^F=(2n+1)\pi/\beta$,
with $n\in \text{Z}$, in contrast with bosonic frequencies $\omega_n^B=2n\pi/\beta$.

\subsection{Components Evaluation}

Let us consider the partition function in components, generalizing the procedure of the pure bosonic model discussed before.
The partition function is then given by the functional integration over all fields presents in the Lagrangian
taking into account the constraints (\ref{c3}) and (\ref{c4}):
\begin{equation}
Z\!=\!\!\int\!\! \mathcal{D}S_{\bf r}\mathcal{D}\psi_{\bf r}\mathcal{D}\bar\psi_{\bf r}\delta\big{(}\sum_{\bf r} S_{\bf r}^2\!-\!N\big{)}
\delta\big{(}\sum_{\bf r}\bar{\psi}_{\bf r}S_{\bf r}\big{)}\delta\big{(}\sum_{\bf r}\psi_{\bf r}S_{\bf r}\big{)}
\delta\big{(}\sum_{\bf r}\bar{\psi}_{\bf r}{\psi}_{\bf r}\!+\!\sum_{{\bf r},{\bf r}^{\prime}}U_{{\bf r},{\bf r}^{\prime}}S_{\bf r}S_{{\bf r}^{\prime}}\big{)}
e^{-\int_{0}^{\beta}d\tau\mathcal{L}_{E}},
\end{equation}
where $\mathcal{L}_{E}$ is the Euclidean version of (\ref{c0}),
\begin{equation}
\mathcal{L}_{Susy}=\frac{1}{2g} \sum_{\bf r}\left( \frac{\partial S_{\bf r}}{\partial\tau}\right)^2
+\frac12\sum_{{\bf r},{\bf r}^{\prime}}J_{{\bf r},{\bf r}^{\prime}}S_{{\bf r}}S_{{\bf r}^{\prime}}+
\frac{1}{\sqrt{g}}\sum_{\bf r}\bar{\psi}_{\bf r}\frac{\partial {\psi}_{\bf r}}{\partial\tau}+
\sum_{{\bf r},{\bf r}^{\prime}}U_{{\bf r},{\bf r}^{\prime}}\bar{\psi}_{{\bf r}}\psi_{{\bf r}^{\prime}}.
\end{equation}
As before, it is convenient to use the integral representation for the delta functions:
\begin{equation}
\delta\big{(}\sum_{\bf r}S_{\bf r}^2-N\big{)}=\int\mathcal{D} \lambda\, e^{{-\int_{0}^{\beta}}d\tau\lambda\big{(}\sum_{\bf r}S_{\bf r}^2-N\big{)}},
\end{equation}
\begin{equation}
\delta\big{(}\sum_{\bf r}\bar{\psi}_{\bf r}S_{\bf r}\big{)}=\int \mathcal{D}\zeta \,e^{-\int_{0}^{\beta}d\tau\sum_{\bf r}\bar\psi_{\bf r}S_{\bf r}\zeta},
\end{equation}
\begin{equation}
\delta\big{(}\sum_{\bf r}\psi_{\bf r}S_{\bf r}\big{)}=\int \mathcal{D}\bar\zeta\, e^{-\int_{0}^{\beta}d\tau\sum_{\bf r}\bar\zeta\psi_{\bf r}S_{\bf r}}
\end{equation}
and
\begin{equation}
\delta\big{(}\sum_{\bf r}\bar{\psi}_{\bf r}{\psi}_{\bf r}+\sum_{{\bf r},{\bf r}^{\prime}}U_{{\bf r},{\bf r}^{\prime}}S_{\bf r}S_{{\bf r}^{\prime}}\big{)}=\int\mathcal{D} \gamma\, e^{{-\int_{0}^{\beta}}d\tau\gamma
\big{(} \sum_{\bf r}\bar{\psi}_{\bf r}{\psi}_{\bf r}+\sum_{{\bf r},{\bf r}^{\prime}}U_{{\bf r},{\bf r}^{\prime}}S_{\bf r}S_{{\bf r}^{\prime}} \big{)}}.
\end{equation}
In this way, the integration over $S_{\bf r}$ and $(\psi_{\bf r},\bar\psi_{\bf r})$ becomes Gaussian and can be performed.
Firstly, let us concentrate on the $S_{\bf r}$ integration by considering the effective partition function
\begin{eqnarray}
Z_{eff}^{\varphi}&\equiv& \int \mathcal{D}S_{\bf r}\,\exp\left\{-\int_{0}^{\beta}d\tau\left[\frac{1}{2g} \sum_{\bf r}\left( \frac{\partial S_{\bf r}(\tau)}{\partial\tau}\right)^2
+\frac12\sum_{{\bf r},{\bf r}^{\prime}}J_{{\bf r},{\bf r}^{\prime}}S_{{\bf r}}S_{{\bf r}^{\prime}}\right.\right.\nonumber\\&+&\left.\left.
\lambda\sum_{\bf r}S_{\bf r}^2+\gamma\sum_{{\bf r},{\bf r}^{\prime}}U_{{\bf r},{\bf r}^{\prime}}S_{\bf r}S_{{\bf r}^{\prime}} +
\sum_{\bf r}\varphi_{\bf r}S_{\bf r} \right]\right\},
\end{eqnarray}
where we defined the real bosonic field $\varphi_{\bf r}\equiv \bar\psi_{\bf r}\zeta+\bar\zeta\psi_{\bf r}$.
Introducing the Fourier transformation of the fields involved and performing the Gaussian integration, we find
\begin{equation}
Z_{eff}^{\varphi}=\exp\left\{-\frac12\text{Tr}\sum_{\bf q}\ln \mathcal{O}_{\bf q}\right\}\exp\left\{-\frac12\int_{0}^{\beta}d\tau\, \sum_{\bf q}\bar\psi_{\bf q}\zeta \mathcal{O}_{\bf q}^{-1}\bar\zeta\psi_{\bf q}\right\},
\end{equation}
with
\begin{equation}
\mathcal{O}_{\bf q}\equiv -\frac{1}{2g}\frac{\partial^2}{\partial\tau^2}+\frac12 J({\bf q})+\lambda +\gamma U({\bf q}).
\end{equation}
Now, we move on to the fermionic integrals by defining another effective partition function whose action is
written in terms of Fourier transformations
\begin{equation}
Z_{eff}\equiv \int \mathcal{D}\bar\psi_{\bf q}\mathcal{D}\psi_{\bf q}\,\exp\left\{-\int_{0}^{\beta}d\tau
\left[\sum_{\bf q}\bar\psi_{\bf q}\left(\frac{1}{\sqrt{g}}
\frac{\partial}{\partial\tau}
+U({\bf q})+\gamma+\frac12\zeta \mathcal{O}_{\bf q}^{-1}\bar\zeta\right)\psi_{\bf q}\right]\right\},
\end{equation}
which gives
\begin{equation}
Z_{eff}=\exp\left[\text{Tr}\sum_{\bf q}\ln\left(-\frac{1}{\sqrt{g}}\frac{\partial}{\partial\tau}-U({\bf q})-\gamma-
\frac12\,\zeta\mathcal{O}_{\bf q}^{-1}\bar\zeta\right)\right].
\end{equation}
So, by putting all together, we get
\begin{equation}
Z=\int\mathcal{D}\lambda\mathcal{D}\gamma\mathcal{D}\bar\zeta\mathcal{D}\zeta e^{-NS_{eff}},
\end{equation}
with
\begin{eqnarray}
S_{eff}&\equiv& -\int_{0}^{\beta}d\tau\lambda +\frac{1}{2N}\text{Tr}\sum_{\bf q}\ln \left(-\frac{1}{2g}\frac{\partial^2}{\partial\tau^2}+\frac12 J({\bf q})+\lambda +\gamma U({\bf q})\right)\nonumber\\&-&
\frac{1}{N}\text{Tr}\sum_{\bf q}\ln\left(-\frac{1}{\sqrt{g}}\frac{\partial}{\partial\tau}-U({\bf q})-\gamma-
\frac12\,\zeta\mathcal{O}_{\bf q}^{-1}\bar\zeta\right).
\end{eqnarray}
Proceeding as before, we use the saddle point method, which in this case implies into the following four conditions
\begin{equation}
\frac{\delta S_{eff}}{\delta \lambda(\tau)}=\frac{\delta S_{eff}}{\delta \gamma(\tau)}=\frac{\delta S_{eff}}{\delta \zeta(\tau)}=
\frac{\delta S_{eff}}{\delta \bar\zeta(\tau)}=0.
\end{equation}
We shall look now for solutions of this equations with all parameters $(\lambda, \gamma, \bar\zeta, \zeta)$ time independent.
The conditions of extremum with respect to the fermionic parameters $\zeta$ and $\bar\zeta$, can be immediately satisfied
if $\zeta=\bar\zeta=0$. So, there are still two remaining conditions. Let us define the corresponding
saddle points as $\lambda\equiv\mu$ and $\gamma\equiv\alpha$. The condition of extremum with respect to the
$\lambda$ can be worked out in the same way that led us to the result (\ref{v1}):
\begin{equation}
1-\frac{1}{N}\sum_{\bf q}\frac{g}{2\omega_{\bf q}^B}\coth\left(\frac{\beta\omega_{\bf q}^B}{2}\right)=0,
\label{sv1}
\end{equation}
but, with the bosonic frequency slightly different, namely, $(\omega_{\bf q}^B)^2\equiv 2g(\mu+\alpha U({\bf q})+\frac{J({\bf q})}{2})$.
Finally, the last condition furnishes
\begin{equation}
\frac{1}{N}\sum_{\bf q}\frac{g}{2\omega_{\bf q}^B}\,U({\bf q})\coth\left(\frac{\beta\omega_{\bf q}^B}{2}\right)-
\frac{1}{N}\sum_{\bf q}\frac{g}{2\omega_{\bf q}^F}\,(U({\bf q})+\alpha)\tanh\left(\frac{\beta\omega_{\bf q}^F}{2}\right)=0,
\label{sv2}
\end{equation}
where the fermionic frequency is
$(\omega_{\bf q}^F)^2\equiv 2g(\frac{\alpha^2}{2}+\alpha U({\bf q})+\frac{J({\bf q})}{2})$. To obtain
this result, we have used the identity
\begin{equation}
\sum_{n=-\infty}^{\infty}\frac{1}{(2n+1)^2+y^2}=\frac{\pi}{2y}\,\tanh\left(\frac{\pi y}{2}\right),
\end{equation}
in the calculation of the trace of the fermionic part, which involves a sum on the fermionic Matsubara
frequencies $\omega_n^F=(2n+1)\pi/\beta$, $n\in \text{Z}$.
The critical behavior of the model can be determined by analyzing these two saddle point conditions near the
critical point, what will be done in the next section.

One last noteworthy remark before closing this section is that the procedure above described can
be also generalized to the superspace, i.e., directly in terms of the superfields. We will not consider this approach in this work.


\section{Critical Behavior}\label{cb}

In order to investigate the critical behavior we need to consider the system near the critical point,
when $\mu\rightarrow 0$ and $\alpha\rightarrow 0$, and the integrals are dominated by the small momenta contributions.
Thus, as mentioned before, we can parameterize the interactions for small values of $|{\bf q}|\equiv q$ as $J({\bf q})\sim q^x$ and $U({\bf q})\sim q^{\frac{x}{2}}$,
respecting the supersymmetry requiring $[U({\bf q})]^2=J({\bf q})$.  Next, we will analyze the critical behavior at
finite temperature as well as at zero temperature, since the system exhibits different behaviors in these two cases.

\subsection{Finite Temperature}

As already mentioned, the supersymmetry is incompatible with temperature, i.e., at finite temperature the
supersymmetry is broken. In this situation, the thermal fluctuations are present and are
responsible to drive the phase transition. At the critical point, the equations (\ref{sv1}) and (\ref{sv2}) become
\begin{equation}
1-\frac{1}{N}\sum_{\bf q}\frac{g_c}{2\sqrt{g_c J({\bf q})}}\coth\left(\frac{\beta\sqrt{g_c J({\bf q})}}{2}\right)=0
\label{sv1a}
\end{equation}
and
\begin{equation}
\frac{1}{N}\sum_{\bf q}\frac{g_c}{2\sqrt{g_c J({\bf q})}}\,U({\bf q})\coth\left(\frac{\beta\sqrt{g_c J({\bf q})}}{2}\right)-
\frac{1}{N}\sum_{\bf q}\frac{g_c}{2\sqrt{g_c J({\bf q})}}\,U({\bf q})\tanh\left(\frac{\beta\sqrt{g_c J({\bf q})}}{2}\right)=0,
\label{sv2a}
\end{equation}
respectively. These integrals converge if $D>x$, what defines the lower critical dimension.

We may determine the critical behavior of the system by subtracting the expression (\ref{sv1}) near the
critical point from (\ref{sv1a}). Technically, near the critical point we can expand the hyperbolic
functions $\coth$ and $\tanh$ for small values of the argument, according to what has been discussed above.

From now, we are going to investigate some simple cases that exhibit interesting critical behaviors:
{\it 1.} $\alpha=0$ with  finite $\mu$ near the critical point; {\it 2.}
$\mu=0$ with finite $\alpha$ near the critical point.

\subsubsection{$\alpha=0$ and finite $\mu$}

By subtracting the equation (\ref{sv1}) (with $\alpha=0$ and small $\mu$)
from (\ref{sv1a}), and evaluating the sum over the momenta, we obtain the following behavior
\begin{equation}
t_g\,\,\,\sim\,\,\,\left\{
\begin{array}
[c]{cc}%
\mu^{\frac{D-x}{x}} & ~~ \left(  D<2x\right)  \\
\mu\ln\mu  & ~~\left( D=2x\right)\\
\mu &~~\left( D>2x\right)
\end{array}
\right.,
\end{equation}
where $t_g\equiv (g-g_c)/g_c$, as in \cite{Vojta1}. Equivalently, of course, we could consider the distance from the critical point
given in terms of the temperature, $t_T\equiv (T-T_c)/T_c$. Clearly, the critical dimension is given by $D_c=2x$.
This is exactly the same result of the purely bosonic spherical model of the section \ref{section3}. We interpret this behavior as the decoupling of the bosonic and fermionic
degrees of freedom when the saddle point value of the parameter $\gamma$ is zero ($\alpha=0$). In this situation, we end up just with
the bosonic model subject to the spherical constraint.


\subsubsection{$\mu=0$ and finite $\alpha$}

For $\mu=0$ and small $\alpha$, after subtracting the equation (\ref{sv1})
from (\ref{sv1a}) and then evaluating the remain sum over the momenta, we find
\begin{equation}
t_g\,\,\,\sim\,\,\,\left\{
\begin{array}
[c]{cc}%
\alpha^{\frac{2(D-x)}{x}} & ~~ \left(  D<\frac{3x}{2}\right)  \\
\alpha\ln\alpha  & ~~\left( D=\frac{3x}{2}\right)\\
\alpha &~~\left( D>\frac{3x}{2}\right)
\end{array}
\right..
\label{q}
\end{equation}
Here, we no longer have the decoupling of the bosonic and fermionic
degrees of freedom, since the correspondent constraint effectively introduces an interaction
mediated by the parameter $\gamma$, whose saddle point value $\alpha$ is now different from zero. Consequently,
the model exhibits a distinct critical behavior and a critical dimension $D_c=3x/2$.

It should be mentioned here, despite the break of the supersymmetry because of the temperature, we could try to
investigate the case when the frequencies become equal, namely, when $\mu=\alpha^2/2$ (in the case of zero temperature
this would correspond the supersymmetric case). The behavior arising from this choice is exactly  that of (\ref{q}).
This can be understood due to the dominance of the term $q^{\frac{x}{2}}$ over $q^{x}$ at small momenta, whenever $\alpha$ is non-vanishing.

\subsection{Zero Temperature}

In the case of zero temperature, $\beta\rightarrow \infty$, the hyperbolic functions  $\coth$ and $\tanh$ become
identically one, and the equations (\ref{sv1}) and (\ref{sv2}) are given by
\begin{equation}
1-\frac{1}{N}\sum_{\bf q}\frac{g}{2\omega_{\bf q}^B}=0
\label{sv10}
\end{equation}
and
\begin{equation}
\frac{1}{N}\sum_{\bf q}\frac{g}{2\omega_{\bf q}^B}\,U({\bf q})-
\frac{1}{N}\sum_{\bf q}\frac{g}{2\omega_{\bf q}^F}\,(U({\bf q})+\alpha)=0.
\label{sv20}
\end{equation}
The integrals converge if $D>x/2$, what defines the lower critical dimension in the quantum case.
The procedure for determining the critical behavior is the
same of the finite temperature. Note that at the critical point $(\mu=\alpha=0)$ the equation (\ref{sv20}) is identically satisfied.

\subsubsection{Supersymmetric case: $\mu=\frac{\alpha^2}{2}$}

The supersymmetric situation is characterized by the equality between bosonic and fermionic frequencies,
$\omega_{\bf q}^B=\omega_{\bf q}^F\equiv \omega_{\bf q}$, which can be achieved by choosing $\mu=\frac{\alpha^2}{2}$, as we said.
The consequence of this choice is that even outside (near) of the critical point we get $\alpha=0$ and also $\mu=0$.
This result indicates the absence of the critical behavior, in agreement with the argument presented in the
Introduction.  It has a simple interpretation: the bosonic and fermionic quantum fluctuations necessary
to drive the quantum phase transition are canceled between themselves in the supersymmetric situation.

\subsubsection{Broken supersymmetry case: $\alpha=0$ and finite $\mu$}

In this situation the frequencies are no longer equal and the supersymmetry is broken. By proceeding as before,
we may determine the quantum critical behavior
\begin{equation}
t_g\,\,\,\sim\,\,\,\left\{
\begin{array}
[c]{cc}%
\mu^{\frac{2D-x}{2x}} & ~~ \left(  D<\frac{3x}{2}\right)  \\
\mu\ln\mu  & ~~\left( D=\frac{3x}{2}\right)\\
\mu &~~\left( D>\frac{3x}{2}\right)
\end{array}
\label{q0}
\right.,
\end{equation}
where now $t_g\equiv (g-g_{c}^0)/g_c^0$, with $g_c^0$ the critical value of $g$ at zero temperature.
The critical dimension is $D_c=3x/2$.
As already argued in the finite temperature case, this result also is the same of that
obtained in the purely bosonic spherical model.

\subsubsection{Broken supersymmetry case: $\mu=0$ and finite $\alpha$}

In this last case, where the supersymmetry remains broken, we get the
following critical behavior
\begin{equation}
t_g\,\,\,\sim\,\,\,\left\{
\begin{array}
[c]{cc}%
\alpha^{\frac{2D-x}{x}} & ~~ \left(  D<x\right)  \\
\alpha\ln\alpha  & ~~\left( D=x\right)\\
\alpha &~~\left( D>x\right)
\end{array}
\right.,
\end{equation}
which, in virtue of the coupling between the bosonic and fermionic degrees of freedom,
it is different from (\ref{q0}), and reveals the critical dimension $D_c=x$.

\section{Conclusions}\label{conclusions}

In this work, we constructed a consistent supersymmetric extension of the quantum spherical model, by
considering the components and the superspace formulations. Afterwards, we calculated the partition function through the
imaginary time formalism, yielding to the saddle point conditions. From these, we studied the critical behavior of the model
in some simple cases for finite and also for zero temperature and determined the critical dimensions.
In general, the model exhibits a critical behavior whenever the supersymmetry is broken, either by thermal effects (classical critical behavior) or
because of the inequality  between the bosonic and fermionic frequencies (quantum critical behavior).

Specifically, we verified that when $\mu$ is finite and $\alpha=0$, the critical behavior ($T\neq 0$ and $T=0$)
is the same of the pure bosonic quantum spherical model, namely, $D_c=2x$ for $T\neq 0$ and $D_c=3x/2$ for $T=0$, which agrees
with the results of \cite{Vojta1}.
In the other case, when $\alpha$ is finite and $\mu=0$,
we obtained a reduction of the critical dimension values: $D_c=3x/2$ for $T\neq 0$ and $D_c=x$ for $T=0$.
When $\mu=\alpha^2/2$, we have a supersymmetric situation at zero temperature. In this case,
the model does not exhibit a quantum phase transition because of the vanishing of the ground state energy
and consequently the absence of the quantum fluctuations.

As final remarks, we stress that some further issues can still be analyzed, for example, the
calculation of thermodynamic quantities. Moreover, the supersymmetric model is an interesting laboratory to investigate some
typical ingredients of the statistical models, such as competing and disorder interactions, including finite-size effects, the
presence of random fields, the existence of quantum Lifshitz points, and their possible influences on the critical properties. Lastly, it could be worth to explore the connection with the supersymmetric field theoretic non-linear sigma model
\cite{Witten,Vecchia,Alvarez}, according to the Stanley's ideas \cite{Stanley}.

\section{Acknowledgments}

The authors thank Prof. S. R. Salinas for reading the manuscript, very useful discussions and criticism.
This work was partially supported by  Conselho
Nacional de Desenvolvimento Cient\'{\i}fico e Tecnol\'ogico (CNPq), Coordena\c c\~ao de Aperfei\c coamento
de Pessoal de N\'ivel Superior (CAPES) and Funda\c{c}\~ao de Amparo a Pesquisa do Estado de S\~ao Paulo (FAPESP).

\appendix

\section{Canonical Quantization}

In this appendix, we consider the supersymmetric mean spherical model, with the constraints imposed as thermal averages,
corresponding to a grand canonical ensemble. From the beginning, in order to simplify the analysis we will
invoke the law of large numbers in the thermodynamic limit, what avoids the presence of orthogonality constraints in the Lagrangian
because of the vanishing of the mixed two point functions
\begin{equation}
\frac{1}{N}\sum_{\bf r}\psi_{\bf r}S_{\bf r}\rightarrow \langle \psi_{\bf r}S_{\bf r} \rangle=0~~~\text{and}~~~
\frac{1}{N}\sum_{\bf r}\bar\psi_{\bf r}S_{\bf r}\rightarrow \langle \bar\psi_{\bf r}S_{\bf r} \rangle=0.
\end{equation}
In the path integral approach, these conditions correspond to the saddle point  choice $\bar\zeta=\zeta=0$.
Without this simplification the diagonalization procedure will involve a transformation of coordinates such that
$A x^2+Bxy+Cy^2\rightarrow \bar A\bar{x}^2+\bar B \bar{y}^2+\bar C$, where $x$ and $y$ generically represent  bosonic and fermionic coordinates.

Next, for our proposals it is convenient to write the Lagrangian (\ref{c0}) in a more symmetric way, already taking into account the Lagrange multipliers
\begin{eqnarray}
\mathcal{L}&=&\frac{1}{2g} \sum_{\bf r} \dot{S}_{\bf r}^2
-\frac12\sum_{{\bf r},{\bf r}^{\prime}}J_{{\bf r},{\bf r}^{\prime}}S_{{\bf r}}S_{{\bf r}^{\prime}}+
\frac{i}{\sqrt{g}}\sum_{\bf r}\bar{\psi}_{\bf r}\dot{\psi}_{\bf r}-
\frac12\sum_{{\bf r},{\bf r}^{\prime}}U_{{\bf r},{\bf r}^{\prime}}[\bar{\psi}_{{\bf r}},\psi_{{\bf r}^{\prime}}]\nonumber\\
&-&\mu\Big{(}\sum_{\bf r}S_{\bf r}^2-N \Big{)}-\alpha\Big{(}\frac12\sum_{\bf r}[\bar{\psi}_{\bf r},{\psi}_{\bf r}]+\sum_{{\bf r},{\bf r}^{\prime}}U_{{\bf r},{\bf r}^{\prime}}S_{\bf r}S_{{\bf r}^{\prime}}\Big{)}-h\sum_{\bf r}S_{\bf r}-
\sum_{\bf r}(\bar\psi_{\bf r} \eta+\bar\eta\psi_{\bf r}),
\end{eqnarray}
where $h$ and $(\bar\eta,\eta)$ are bosonic and fermionic external fields.
By introducing the momenta
\begin{equation}
P_{\bf r}=\frac{\partial \mathcal{L}}{\partial \dot S_{\bf r}},~~~\Pi_{\bf r}=\frac{\partial \mathcal{L}}{\partial \dot\psi_{\bf r}}~~~
\text{and}~~~\bar{\Pi}_{\bf r}=\frac{\partial \mathcal{L}}{\partial \dot{\bar\psi}_{\bf r}},
\end{equation}
we can easily determine the Hamiltonian by means of a Legendre transformation
\begin{eqnarray}
\mathcal{H}&=&\frac{g}{2} \sum_{\bf r} P_{\bf r}^2
+\frac12\sum_{{\bf r},{\bf r}^{\prime}}J_{{\bf r},{\bf r}^{\prime}}S_{{\bf r}}S_{{\bf r}^{\prime}}+
\frac12\sum_{{\bf r},{\bf r}^{\prime}}U_{{\bf r},{\bf r}^{\prime}}[\bar{\psi}_{{\bf r}},\psi_{{\bf r}^{\prime}}]+
\mu\Big{(}\sum_{\bf r}S_{\bf r}^2-N \Big{)}\nonumber\\
&+&\alpha\Big{(}\frac12\sum_{\bf r}[\bar{\psi}_{\bf r},{\psi}_{\bf r}]+\sum_{{\bf r},{\bf r}^{\prime}}U_{{\bf r},{\bf r}^{\prime}}S_{\bf r}S_{{\bf r}^{\prime}}\Big{)}+
h\sum_{\bf r}S_{\bf r}+\sum_{\bf r}(\bar\psi_{\bf r} \eta+\bar\eta\psi_{\bf r}).
\end{eqnarray}
From now on, we interpret the variables as quantum operators. After writing the Hamiltonian in the Fourier space and then
performing the standard Bogoliubov transformations, we can find the diagonal form in terms of creation and annihilation operators
\begin{equation}
\mathcal{H}=\sum_{\bf q}\omega_{\bf q}^B\left(a_{\bf q}^{\dagger}a_{\bf q}+\frac12\right)+
\sum_{\bf q}\omega_{\bf q}^F\left(c_{\bf q}^{\dagger}c_{\bf q}-\frac12\right)-\mu N-\frac{N h^2}{4\mu}-\frac{N\bar\eta\eta}{\alpha},
\end{equation}
with the frequencies $(\omega_{\bf q}^B)^2\equiv 2g(\mu+\alpha U({\bf q})+\frac{J({\bf q})}{2})$ and
$(\omega_{\bf q}^F)^2\equiv 2g(\frac{\alpha^2}{2}+\alpha U({\bf q})+\frac{J({\bf q})}{2})$ and the bosonic and fermionic
operators  satisfying
\begin{equation}
[a_{\bf q},a_{{\bf q}^{\prime}}]=0,~~~[a_{\bf q}^{\dagger},a_{{\bf q}^{\prime}}^{\dagger}]=0,~~~
[a_{\bf q},a_{{\bf q}^{\prime}}^{\dagger}]=\delta_{{\bf q},{\bf q}^{\prime}}
\end{equation}
and
\begin{equation}
\{c_{\bf q},c_{{\bf q}^{\prime}}\}=0,~~~\{c_{\bf q}^{\dagger},c_{{\bf q}^{\prime}}^{\dagger}\}=0,~~~
\{c_{\bf q},c_{{\bf q}^{\prime}}^{\dagger}\}=\delta_{{\bf q},{\bf q}^{\prime}}.
\end{equation}
The partition function can be straightforwardly evaluated according to $Z=\text{Tr}\,e^{-\beta\mathcal{H}}$ and the
connection with the thermodynamics is through the free energy
\begin{equation}
f=-\frac{1}{\beta N}\ln Z=-\frac{1}{N\beta}\sum_{\bf q}\left[\ln\cosh\left(\frac{\beta\omega_{\bf q}^F}{2}\right)-
\ln\sinh\left(\frac{\beta\omega_{\bf q}^B}{2}\right)\right]-\mu-\frac{ h^2}{4\mu}-\frac{\bar\eta\eta}{\alpha}.
\end{equation}
Finally, the constraints are implemented as
\begin{equation}
\frac{\partial f}{\partial \mu}=0~~~\text{and}~~~\frac{\partial f}{\partial \alpha}=0.
\end{equation}
The resulting conditions coincide with equations (\ref{sv1}) and (\ref{sv2}), when $h=0$ and $\bar\eta=\eta=0$.


\end{document}